\pdfoutput=1
\documentclass[aps,prd,amsmath,floats,floatfix, twocolumn,
superscriptaddress,nofootinbib,showpacs,longbibliography]{revtex4-1}
\usepackage[T1]{fontenc}
\usepackage[utf8]{inputenc}
\usepackage{lmodern}
\usepackage{verbatim}
\usepackage[dvipsnames, usenames]{xcolor}
\definecolor{linkcolor}{rgb}{0.0,0.3,0.5}
\usepackage[hypertexnames=false, unicode, colorlinks=true, linkcolor=linkcolor,
citecolor=linkcolor, filecolor=linkcolor,urlcolor=linkcolor,
pdfusetitle]{hyperref}
\usepackage[all]{hypcap}
\usepackage{graphicx}
\usepackage{xspace}
\usepackage{amssymb}
\usepackage{microtype}

\usepackage[english]{babel}
\usepackage{blindtext}

\usepackage[normalem]{ulem} 
\usepackage{bm} 
\usepackage{mathrsfs}

\DeclareMathAlphabet{\mathpzc}{OT1}{pzc}{m}{it}

\usepackage[nomessages]{fp}

\newcommand{\CITEME}[1]{\textcolor{red}{\textbf{[CITE]}}}

\begin{document}
	
\title{Gravitational waves from black hole emission}

\newcommand{\UMassDMath}{\affiliation{Department of Mathematics,
		University of Massachusetts, Dartmouth, MA 02747, USA}}
\newcommand{\UMassDPhy}{\affiliation{Department of Physics,
		University of Massachusetts, Dartmouth, MA 02747, USA}}
\newcommand{\CSCVRUMass}{\affiliation{Center for Scientific Computing and Data Science Research, University of Massachusetts, Dartmouth, MA 02747, USA}}
\newcommand{\URI}{\affiliation{Department of Physics and Center for Computational Research, University of Rhode Island, Kingston, RI 02881, USA}}  
\newcommand{\TAPIR}{\affiliation{Theoretical Astrophysics, Walter Burke Institute for Theoretical Physics, California Institute of Technology, Pasadena, California 91125, USA}}
\newcommand{\KITP}{\affiliation{Kavli Institute of Theoretical Physics, University of California Santa Barbara, Santa Barbara, CA 93106, USA}}

\author{Tousif Islam}
\email{tousif.islam@kitp.ucsb.edu}
\KITP
\TAPIR
\UMassDPhy
\UMassDMath
\CSCVRUMass

\author{Gaurav Khanna}
\URI
\UMassDPhy
\CSCVRUMass

\author{Steven L. Liebling}
\affiliation{Long Island University, Brookville, New York 11548, USA}

\begin{abstract}
Using adiabatic point-particle black hole perturbation theory, we simulate plausible gravitational wave~(GW) signatures in two exotic scenarios (i)~where a small black hole is emitted by a larger one (`black hole emission') and (ii)~where a small black hole is emitted by a larger one and subsequently absorbed back (`black hole absorption'). While such scenarios are forbidden in general relativity!(GR), alternative theories (such as certain quantum gravity scenarios obeying the weak gravity conjecture, white holes, and Hawking radiation) may allow them. 
By leveraging the phenomenology of black hole emission and absorption signals, we introduce straightforward modifications to existing gravitational waveform models to mimic gravitational radiation associated with these exotic events. We anticipate that these (incomplete but) initial simulations, coupled with the adjusted waveform models, will aid in the development of null tests for GR using GWs.
\end{abstract} 

\maketitle
	
\section{Introduction}
\label{sec:intro}

General relativity (GR) remains the most successful theory of gravity to date, having passed numerous tests at both the solar and cosmological scales, as well as in weak and strong gravity regimes~\cite{Will:2014kxa}. Gravitational wave~(GW) astronomy offers a new opportunity to scrutinize the validity of GR, particularly in the strong gravity regime. To date, analyses of detected GW signals have not revealed any violations of GR~\cite{LIGOScientific:2019fpa,LIGOScientific:2020tif,LIGOScientific:2021sio}. Nonetheless, developing efficient tests to detect potential deviations from GR predictions in GW data remains a significant pursuit within the GW community. 

These tests generally fall into two categories: (i)~theory-specific tests, which assess whether the data supports predictions from alternative theories of gravity (such as Chern-Simons gravity, Gauss-Bonnet theory of gravity, Scalar-Tensor-Vector Gravity)~\cite{Canizares:2012is,Yunes:2010yf,Perkins:2021mhb,Carson:2020ter,Liu:2019cxm} or exotic sources of radiation (such as boson star binaries or cosmic strings)~\cite{Bezares:2022obu,Bezares:2018qwa,Palenzuela:2017kcg,Ahmed:2024cty}, and (ii)~theory-agnostic null tests of GR, which look for potential generic deviations from GR~\cite{Islam:2019dmk,Dhanpal:2018ufk,Krishnendu:2017shb,Johnson-Mcdaniel:2018cdu,Kastha:2018bcr,Carullo:2018gah,Carullo:2019flw,Ghosh:2021mrv,Edelman:2020aqj,Isi:2019asy,Isi:2019aib,Isi:2020tac,CalderonBustillo:2020rmh,Ghosh:2016qgn,Ghosh:2017gfp,Capano:2020dix}. The latter type of tests are often guided by common phenomenological predictions from various alternative theories of gravity. However, in both cases, performing the test on real GW data requires the availability of a fast and accurate waveform model that includes necessary physics. Typically, for alternative theories of gravity, no waveform model includes radiation from inspiral to merger. Instead, existing waveform models in GR can be suitably modified using phenomenological relations (such as the parameterized post-Einsteinian or ppE framework~\cite{Yunes:2009ke}) to mimic waveforms in alternative theories of gravity. Obtaining such phenomenological modifications involves simulating a handful of signals under those specific theories of gravity or for specific binary sources.

One potential violation of GR may involve a black hole transforming (say, via emission, decay, or splitting) into a pair of smaller black holes. A number of unproven scenarios allow such processes, and hence tests that may observe the emission of a small black hole by a larger one are potentially interesting and worthwhile. One such scenario is the possibility that extremal black holes might decay, with consequences for the weak gravity conjecture~\cite{Arkani-Hamed:2006emk,Cheung:2018cwt,McInnes:2021zlt,Marolf_2005,Emparan:2003sy}. Further motivation comes from the exotic possibilities of emission from either white holes~\cite{Gaur:2023ved} or wormholes~\cite{Dent:2020nfa}. Hawking radiation itself represents the emission from black holes of particles, although generally one only considers the low energy option provided by massless photons. 

The idea pursued here is to simulate gravitational waveforms for the emission of a small black hole (the secondary) from a large black hole (the primary). One could imagine doing so with fully nonlinear, numerical relativity~(NR), but  then one confronts the difficulty of creating the conditions for the emission of a small black hole. Reversal of the signal resulting from a binary black hole merger should describe the emission of a black hole (from the merged remnant), but along with the emission would be the ``absorption'' of all the gravitational waves that had been emitted in the merger and being input into the system from infinity and the horizons.

Instead of fully nonlinear solutions, the point-particle black hole perturbation theory~(ppBHPT) solutions of the Teukolsky equation for linearized gravity are more appropriate for this problem. First, such solutions are obtained in two distinct steps with the path of the secondary computed first and the gravitational waves computed second. Hence, one can reverse the path without needing to set up the unphysical absorption of gravitational waves in the computational simulation. Second, that these solutions become more accurate with more extreme mass ratios and therefore are more closely relevant for the scenario of black hole emission.

Here, using adiabatic point-particle black-hole perturbation theory\footnote{By ``adiabatic'' we mean that we work within the approximation that the radiated energy is sufficiently small that the system energy remains constant and orbits are quasi-circular.}, we simulate plausible GWs in two related exotic scenarios: (i) where a small black hole is emitted by a larger one (`black hole emission') and (ii) where a small black hole is emitted by a larger one and subsequently absorbed back (`black hole absorption'). We detail our numerical method in Section~\ref{sec:method}, followed by our results in Section~\ref{sec:results}. By leveraging the phenomenology of black hole emission and absorption signals, we introduce straightforward modifications to existing waveform models to mimic gravitational radiation associated with these exotic events. Finally, we discuss these results in light of future detections in Section~\ref{sec:discussion}.

\section{Numerical Setup}
\label{sec:method}

For the rest of the paper, we adopt natural units $G=c=1$ and work in the center of mass frame of the binary. We use $m_1$ ($m_2$) to denote the mass of the larger (smaller) black hole. We further define the total mass of the system as $M\equiv m_1+m_2$ and the mass ratio as $q\equiv m_1/m_2$. For all of the simulations, we use $q=10^{3}$. 

In the ppBHPT framework, the smaller black hole is modeled as a point-particle with no internal structure and a mass of $m_2$ moving in the spacetime of the larger Kerr black hole with mass $m_1$ and spin angular momentum per unit mass $a$.  The trajectory of the smaller black hole is given by a set of four dynamical variables $\{r, \phi, p_r, p_{\phi} \}$ where $r$ is the radial separation between the two black holes, $\phi$ is the orbital phase, $p_r$ is the radial momentum, and $p_{\phi}$ is the angular momentum. Gravitational radiation 
from such a binary 
is well described by the Teukolsky equation:
\begin{eqnarray}
\label{teuk0}
&&
-\left[\frac{(r^2 + a^2)^2 }{\Delta}-a^2\sin^2\theta\right]
\partial_{tt}\Psi
-\frac{4 m_1 a r}{\Delta}
\partial_{t\phi}\Psi \nonumber \\
&&- 2s\left[r-\frac{m_1(r^2-a^2)}{\Delta}+ia\cos\theta\right]
\partial_t\Psi\nonumber\\  
&&
+\,\Delta^{-s}\partial_r\left(\Delta^{s+1}\partial_r\Psi\right)
+\frac{1}{\sin\theta}\partial_\theta
\left(\sin\theta\partial_\theta\Psi\right)+\nonumber\\
&& \left[\frac{1}{\sin^2\theta}-\frac{a^2}{\Delta}\right] 
\partial_{\phi\phi}\Psi +\, 2s \left[\frac{a (r-m_1)}{\Delta} 
+ \frac{i \cos\theta}{\sin^2\theta}\right] \partial_\phi\Psi  \nonumber\\
&&- \left(s^2 \cot^2\theta - s \right) \Psi = -4\pi\left(r^2+a^2\cos^2\theta\right)T  \,  ,
\end{eqnarray}
sourced by the moving particle, where $\Delta = r^2 - 2 m_1 r + a^2$ and $s$ is the ``spin weight'' of the field. 
The $s=-2$ case for $\Psi$ describes the radiative degrees of freedom of the gravitational field, the Weyl scalar 
$\psi_4$, in the radiation zone, and is directly related to the Weyl curvature scalar as $\Psi = (r -ia\cos\theta)^4\psi_4$. 
The source term $T$ in Eq.~(\ref{teuk0}) for the  smaller compact object $m_2$ 
is related to the energy-momentum tensor $T_{\alpha\beta}$ of a point particle.
The Weyl scalar $\psi_4$ can then be integrated twice at future null infinity ${\mathscr{I}^+}$ to find the two polarization states $h_{+}$ and $h_{\times}$ of the transverse-traceless metric perturbations,
\begin{equation}
\psi_4 = \frac{1}{2}\left(\frac{\,\partial^2h_{+}}{\,\partial t^2}-i\frac{\,\partial^2h_{\times}}{\,\partial t^2}\right)\, .
\end{equation}
The complex gravitational wave strain
\begin{align}
h_{+}(t,\theta, \phi;q) &- {\mathrm i} h_{\times}(t,\theta, \phi; q) \nonumber \\
& =  \sum_{\ell=2}^{\infty} \sum_{m=-\ell}^{\ell} h_{\ell m}(t;q) {}_{-2}Y_{\ell m}(\theta, \phi) \, ,
\end{align}
can be formed from the two polarization states,
which is subsequently decomposed into a basis of spin-weighted spherical harmonics ${}_{-2}Y_{\ell m}$. Each mode can then be decomposed into an amplitude $A_{\ell m}$ and phase $\phi_{\ell m}$ such that: $h_{\ell m}(t;q) := A_{\ell m}(t;q) e^{i \phi_{\ell m}(t;q)}$.

Before solving the Teukolsky equation, we must know the worldline (trajectory) of 
the smaller black hole, such as radiation-reaction driven orbits or geodesic plunge orbits. Once the trajectory of the perturbing compact body is fully specified as described above, 
we solve the inhomogeneous Teukolsky equation in the time domain while feeding the trajectory 
information into the particle source-term of the equation 
This involves a multi-step process: (i) rewriting the Teukolsky equation using compactified hyperboloidal 
coordinates that allow for the extraction of the gravitational waveform directly at null infinity while 
also solving the ``outer boundary problem'' of the finite computational domain; (ii) transforming the 
equation into a set of (2+1) dimensional PDEs by using the axisymmetry of the background Kerr space-time, 
and separating the dependence on azimuthal coordinate; (iii) recasting these equations into a first-order, 
hyperbolic PDE system; and lastly (iv) implementing a high-order WENO (3,5) finite-difference scheme with 
Shu-Osher (3,3) explicit time-stepping. Details of our Teukolsky solver code is provided in Refs.~\cite{Sundararajan:2007jg,Sundararajan:2008zm,Sundararajan:2010sr,Zenginoglu:2011zz,WENO}.

One of the challenges in simulating plausible gravitational radiation from black hole emission is obtaining an accurate trajectory for the emitted black hole. In the absence of any physically motivated and theoretically consistent framework to predict this trajectory, a simpler approach is to reverse the trajectory of a plunging smaller black hole. However, a complication arises: in adiabatic radiation-reaction driven plunging orbits, energy and angular momentum are radiated away. Reversing such orbits would also reverse the corresponding energy and angular momentum emission, which is undesirable.

To address this, we choose a geodesic plunge trajectory that closely approximates a circular orbit for our secondary black hole, and then reverse this geodesic trajectory. Since geodesics do not involve the radiative losses mentioned earlier, and because the reversal of a geodesic is itself a geodesic, we obtain a physically meaningful path. One drawback of this method is that the resulting signal duration is shorter, which limits the signal-to-noise ratio (SNR) of our signals. On the other hand, short plunging geodesics offer good approximations for physical black hole binary systems, especially for large mass-ratios. This is because the gravitational wave emission is weak and is unable to impact the plunge trajectory significantly over its  short duration. 

Similarly, to simulate the scenario of a smaller black hole being emitted from a larger black hole and then being absorbed back, we first reverse a geodesic trajectory to mimic the black hole emission. We then combine this with a geodesic plunge orbit to mimic the absorption process.

\begin{figure*}
\includegraphics[width=\textwidth]{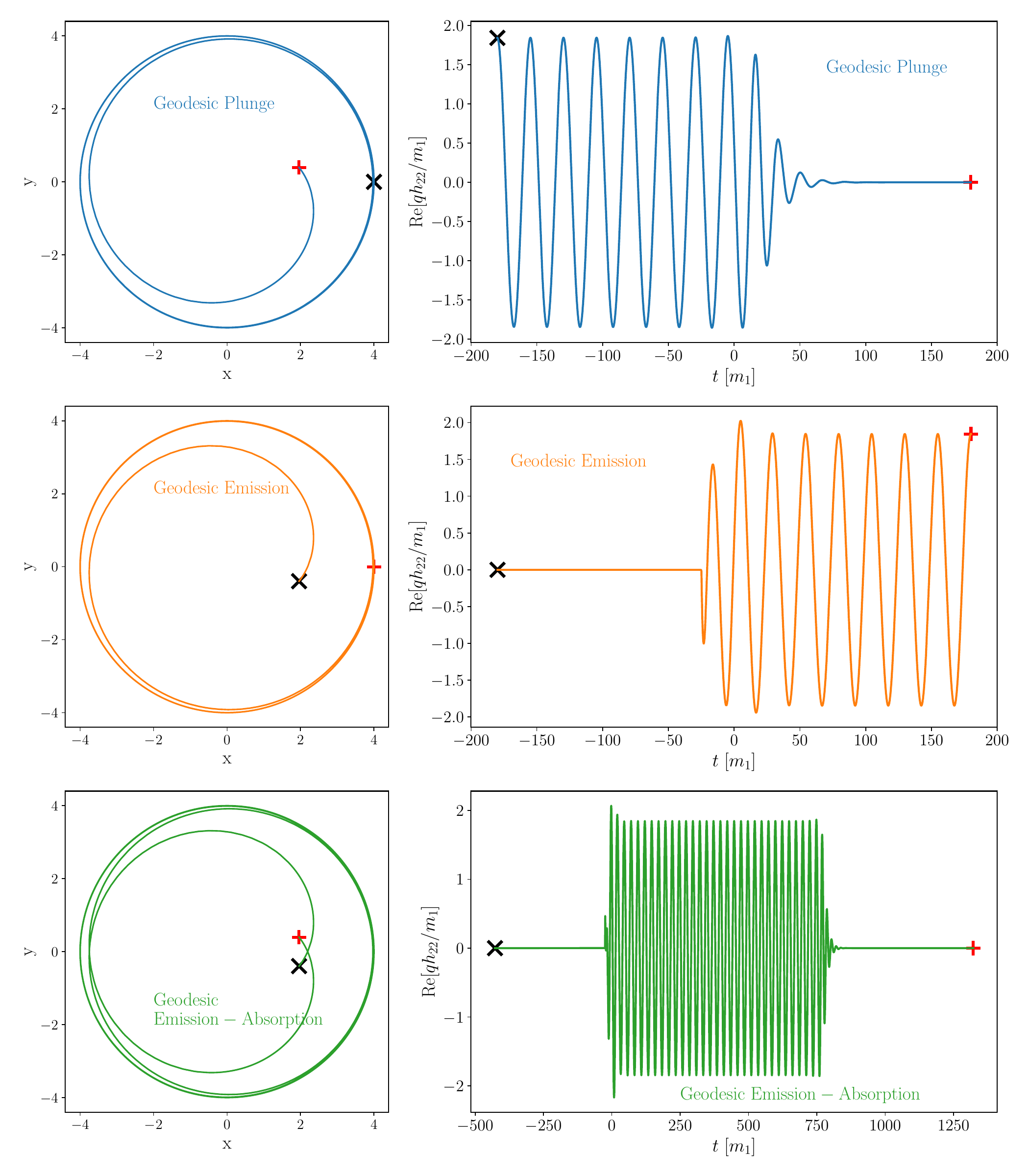}
\caption{\textit{Gravitational waveforms:} \textbf{Top:} The orbit of a conventional geodesic plunge (left panel) and corresponding $(2,2)$ mode gravitational waveform (right panel). \textbf{Middle:} The orbit of a black hole emission event (left panel) and corresponding $(2,2)$ mode gravitational waveform (right panel) \textbf{Bottom:} The orbit of a black hole emission and subsequent absorption (left panel) and corresponding $(2,2)$ mode gravitational waveform (right panel). Black crosses denote the start of the simulation and red squares indicate the end of the simulation. More details can be found in Section \ref{sec:results}.}
\label{fig:waveforms}
\end{figure*}

\begin{figure*}
\includegraphics[width=\textwidth]{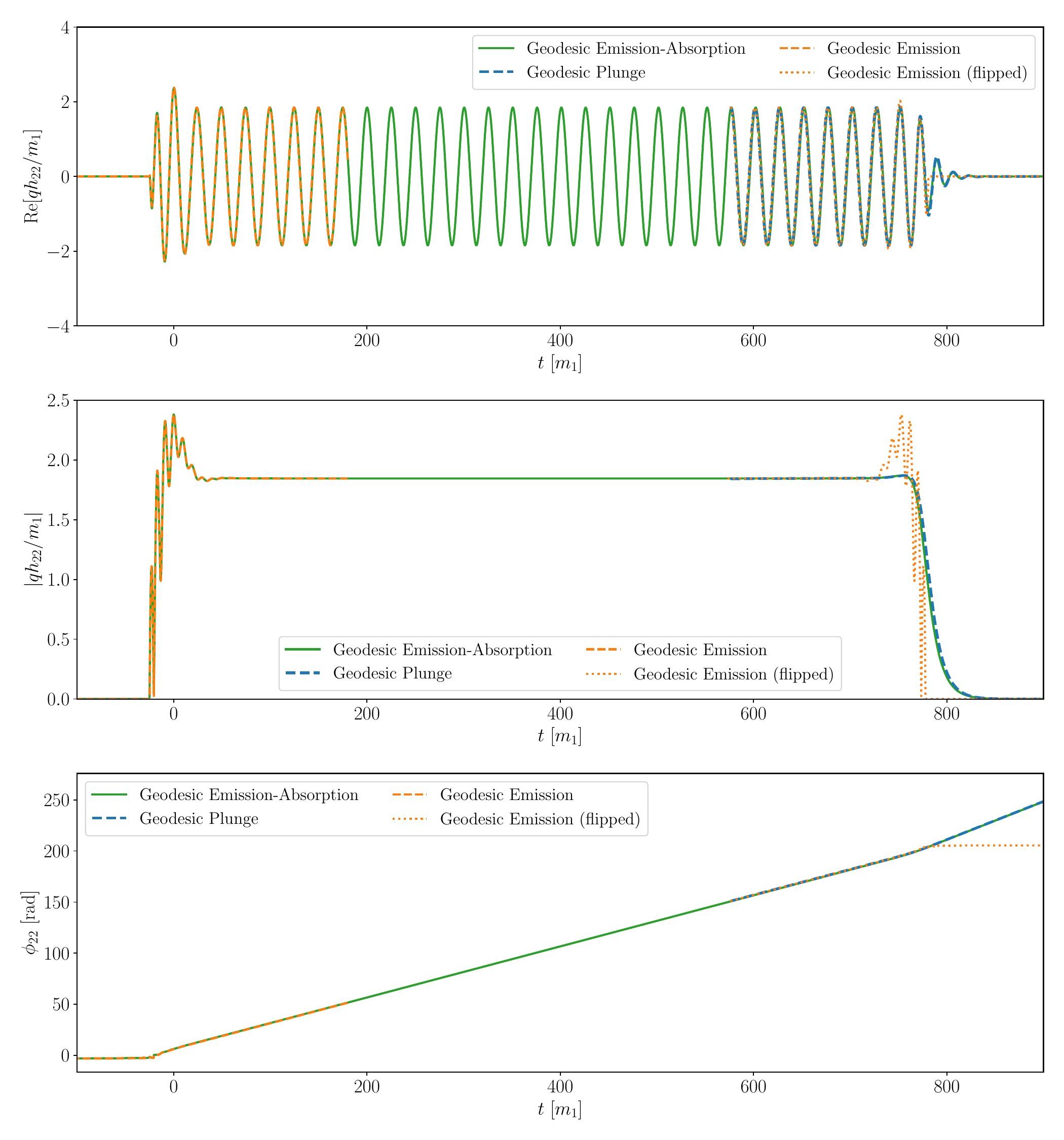}
\caption{\textit{Comparison of waveforms:} \textbf{Top:} The $(2,2)$ mode waveform for the conventional geodesic plunge (solid blue line), geodesic emission (dashed orange line), geodesic emission waveform after the flip (dotted orange line) and geodesic emission-absorption (solid green line). We apply additional time and phase shift to the geodesic plunge and geodesic emission waveform to ensure they match the corresponding portion of the geodesic plunge-absorption waveform. \textbf{Middle:} The amplitudes. \textbf{Bottom:} The phases. More details can be found in Section \ref{sec:results}.}
\label{fig:waveform_comparison}
\end{figure*}

\section{Results}
\label{sec:results}
In this section, we present our main findings, focusing on the morphology of the emission and absorption waveforms and potential implications for detection and data analysis.

We first simulate the gravitational waveform for a merging binary on a geodesic plunge orbit, decomposing the waveform into its constituent harmonic modes.
This simulation starts with an initial energy $E=1$ and angular momentum $p_{\phi}=4 -\epsilon$ where $\epsilon$ is a small number we control to set the length of the plunge. In Figure~\ref{fig:waveforms} (\textit{upper left panel}), we show the geodesic plunge trajectory (as a blue solid line) in the $xy$ plane, where $x=r \cos \phi$ and $y=r\sin \phi$. The corresponding quadrupolar mode (i.e., the $(2,2)$ mode) of the gravitational waveform is presented in Figure~\ref{fig:waveforms} (\textit{upper right panel}). It exhibits a characteristic plunge waveform followed by ringdown. All the waveforms scale with $q$ and $m_1$, while we express time $t$ in units of $m_1$.

Next, following the prescription given in Section~\ref{sec:method}, we construct trajectories for the black hole emission and subsequent absorption using the same geodesic obtained earlier. These trajectories are shown in the middle left panel (solid orange line) and lower left panel (solid green line), respectively, in Figure~\ref{fig:waveforms}. The corresponding quadrupolar mode of gravitational waves is presented in the middle right panel (solid orange line) and lower right panel (solid green line), respectively, in Figure~\ref{fig:waveforms}. The black hole emission trajectory exhibits a reverse chirp behavior, whereas the emission and absorption case shows a reverse chirp followed by the usual chirp (like the geodesic plunge case).

The distinctive characteristic of the black hole emission waveform (Figure~\ref{fig:waveforms}; \textit{middle right panel}) is the radiation commencing only when the secondary is emitted. One could imagine for the exotic case of emission that a black hole about to emit a small part of itself may involve distortions that give rise to a ``precursor signal" so-to-speak. However, our approach does not account for such an effect, and it remains unclear how these effects might manifest.

Next, we ask whether reversing the geodesic simply results in a time reversal of the obtained waveform (in the black hole emission scenario). In other words, how different is the waveform of a geodesic plunge from the black hole emission waveform obtained after flipping it in time (where we transform $t \rightarrow T - t$ with $T$ being a constant)? A similar comparison is also made for the waveform obtained in the black hole emission and subsequent absorption scenario. In Figure~\ref{fig:waveform_comparison}, we show the real part of the $(2,2)$ mode waveform (along with the amplitude and phase) for the conventional geodesic plunge (solid blue line), geodesic emission (dashed orange line), geodesic emission waveform after the flip (dotted orange line) and geodesic emission-absorption (solid green line). We apply additional time and phase shifts to the geodesic plunge and geodesic emission waveforms to ensure that they match the corresponding portions of the geodesic plunge-absorption waveform.

\begin{figure}
\includegraphics[width=\columnwidth]{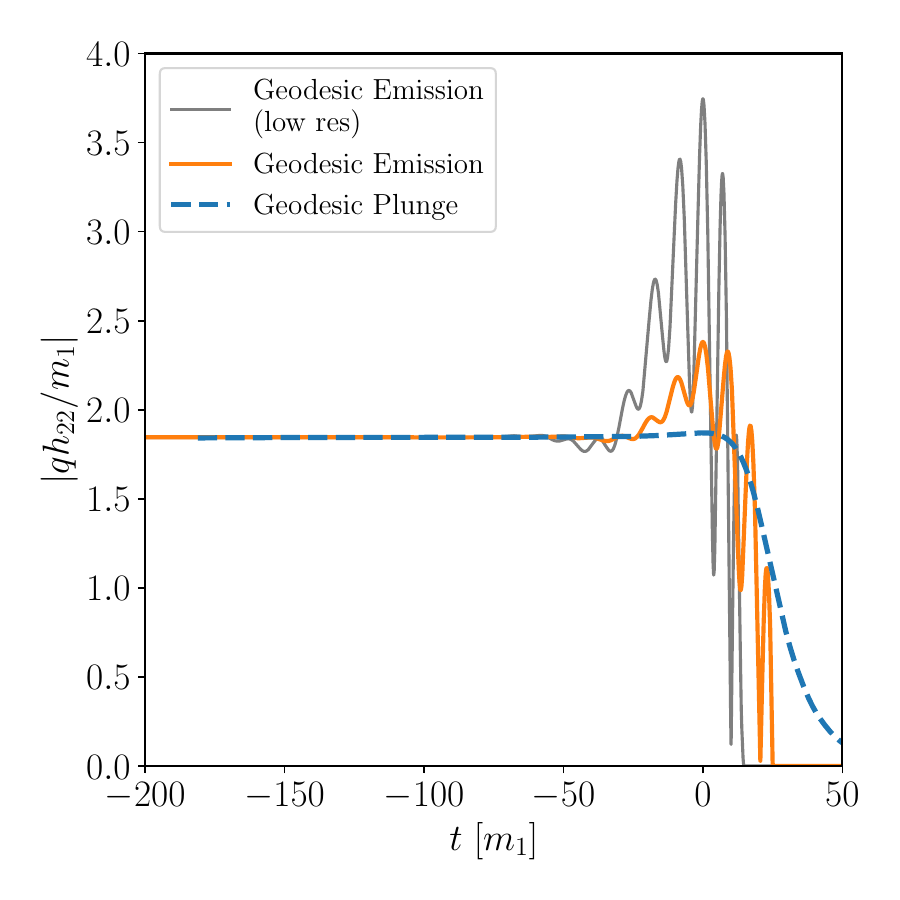}
\caption{\textit{Origin of oscillations in amplitudes:} We show the amplitude of the $(2,2)$ mode waveform obtained from two different simulations with varying resolutions. For the reversed geodesic waveforms, we flip the time so that it matches the plunge geodesic waveforms. These results suggest that at least some part of the oscillatory behavior corresponds to resolution-dependent junk radiation.}
\label{fig:resolution}
\end{figure}

\begin{figure*}
\includegraphics[width=\textwidth]{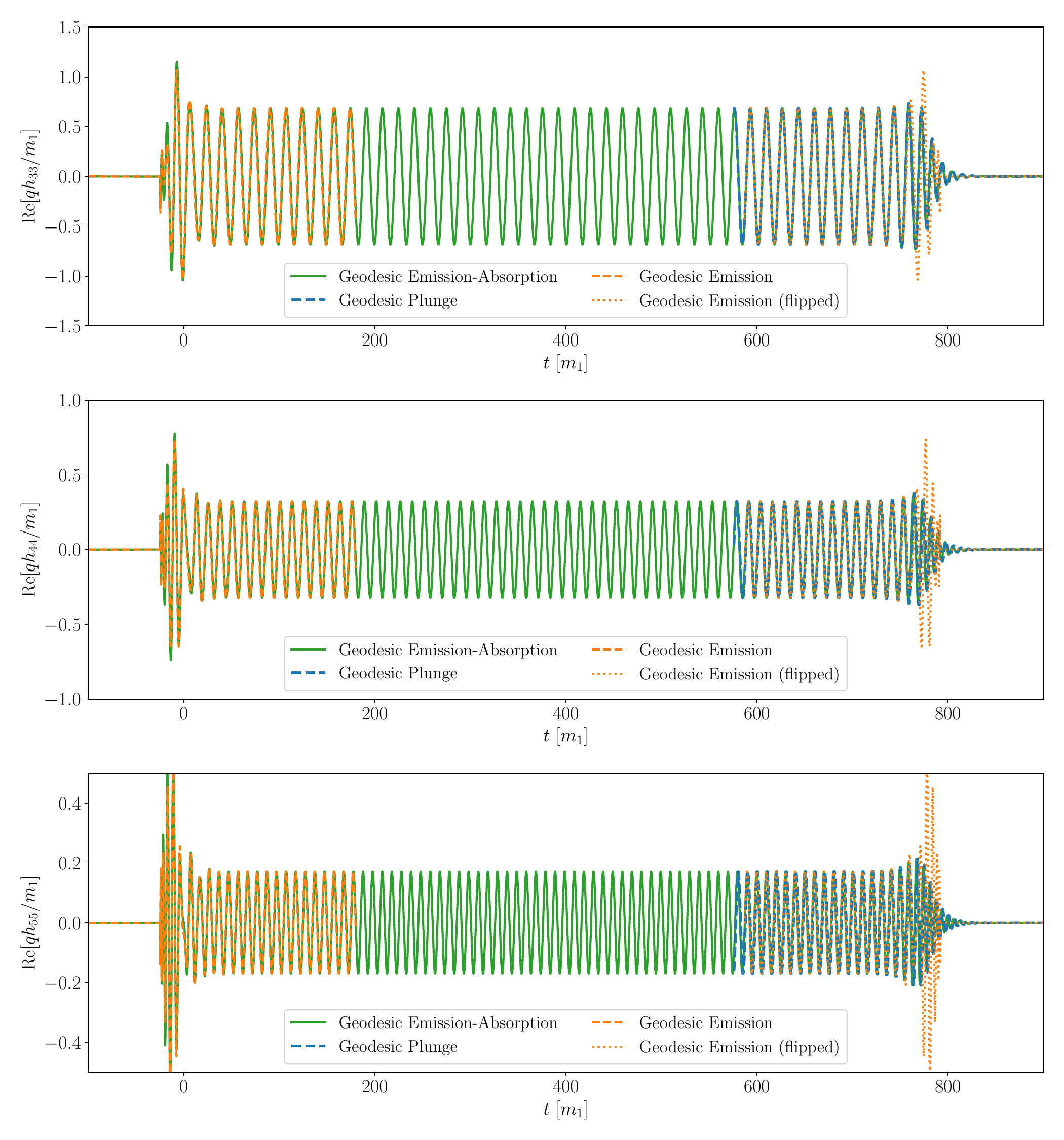}
\caption{\textit{Comparison of higher modes:} The $(3,3)$ (upper panel), $(4,4)$ (middle panel) and $(5,5)$ (lower panel) mode waveforms for the conventional geodesic plunge (solid blue line), geodesic emission (dashed orange line), geodesic emission waveform after the flip (dotted orange line) and geodesic emission-absorption (solid green line). We apply additional time and phase shift to the geodesic plunge and geodesic emission waveform to ensure they match the corresponding portion of the geodesic plunge-absorption waveform. More details can be found in Section \ref{sec:results}.}
\label{fig:mode_comparison}
\end{figure*}

We observe that the emission waveform after flipping is nearly identical to the conventional plunge waveform for the first few cycles. However, as they approach the peak amplitude (i.e. merger), visible differences in both amplitudes (and phases) become apparent at around $30m_1$ before merger.
It is crucial to note that, away from merger, the agreement between the geodesic plunge waveform and the black hole emission waveform is not by construction.  Note that the waveform amplitudes of the forward geodesic plunge waveform and the reversed emission one match well in the circular orbit phase. While this is not unexpected, it does offer a sanity check of sorts on our approach and results. We further note that the waveform from the black hole emission and absorption matches the waveform from the black hole emission scenario very well in the initial part and matches the conventional geodesic plunge waveform in the later part.

We find oscillatory features in the waveform (in particular in the amplitudes) around the peak amplitude.
These oscillatory features are more likely to be associated with the junk radiation at the start of the simulation\footnote{Fully nonlinear numerical simulations of black hole binaries have long confronted the problem of spurious gravitational radiation~\cite{Higginbotham:2019wbx}.}. To understand this, we perform the simulations with different grid resolutions. For the high resolution simulation, we use a grid resolution of $M/160$ whereas the low resolution simulation uses a resolution of $M/80$. Note that, unless otherwise specified, all results shown in this section are from the higher resolution simulation. We find that the higher resolution simulation displays a lower degree of oscillatory features. In Figure~\ref{fig:resolution}, we show the $(2,2)$ mode amplitude of the geodesic emission waveforms for two different resolution simulations.
If all the oscillation corresponds to junk radiation, the reverse plunge waveform otherwise matches the usual plunge waveform very well. 

Next, we focus our attention on higher-order spherical harmonic modes. In Figure~\ref{fig:mode_comparison}, we show the $(3,3)$ (upper panel), $(4,4)$ (middle panel), and $(5,5)$ (lower panel) modes of the gravitational waveform obtained for the conventional geodesic plunge (solid blue line), geodesic emission (dashed orange line), geodesic emission waveform after the flip (dotted orange line), and geodesic emission-absorption (solid green line). Similar to the waveforms shown in Figure~\ref{fig:waveform_comparison}, we apply additional time and phase shifts.

We further note that the oscillatory features in the waveforms around the maximum amplitude become more pronounced for the higher-order modes. This is not unusual, as these oscillations are related to the junk radiation associated with resolution (as demonstrated in Figure~\ref{fig:resolution}), and resolving higher-order modes typically requires a smaller grid spacing compared to the dominant quadrupolar mode.
Except for the portion around the maximum peak (which corresponds to the merger in conventional geodesic plunge orbits), the higher modes in the (flipped) waveforms obtained for the black hole emission scenario agree quite well with the conventional geodesic plunge waveforms.

\section{Implications}
\label{sec:discussion}
Our results in Section~\ref{sec:results} (using geodesic plunge) rely on the assumption that the orbits of the smaller black hole in the emission (and absorption) scenario are similar to conventional plunge orbits (after reversing). In reality, these orbits may differ from typical plunge orbits due to quantum effects associated with black hole emission. However, since these beyond GR corrections are expected to be very small, any differences in orbit structure are likely minor. Furthermore, our findings are based on the use of geodesic orbits for simplicity. Nevertheless, these results provide qualitative insights into gravitational waveforms in black hole emission (and absorption) scenarios. Specifically, they suggest that, in the absence of significant quantum effects, the waveform (after removing any suspected junk radiation portion) from the black hole emission scenario simplifies to a reverse chirp signal.

\subsection{Radiation-reaction driven orbits}
As a potential next step from our geodesic orbit scenarios, we compute a radiation-reaction driven orbit for the smaller black hole in a binary system with mass ratio $q=10^3$ and spin parameter $a=0$. This orbit exhibits three distinct regimes: an initial adiabatic inspiral, a late-stage geodesic plunge into the horizon, and a transitional regime between these phases. 
During the initial adiabatic inspiral, the particle follows a sequence of geodesic orbits driven by radiative energy and angular momentum losses. The flux radiated to null infinity and through the event horizon is computed using the frequency-domain Teukolsky equation \cite{Fujita:2004rb, Fujita:2005kng, Mano:1996vt, throwe2010high} implemented in the open-source code \texttt{GremlinEq} \cite{OSullivan:2014ywd, Drasco:2005kz} from the Black Hole Perturbation Toolkit \cite{BHPToolkit}. The inspiral trajectory is then extended to include a geodesic plunge and a smooth transition region, following a procedure similar to that proposed by Ori-Thorne \cite{Ori:2000zn, Hughes:2019zmt, Apte:2019txp}.

\begin{figure*}[h]
\includegraphics[width=\textwidth]{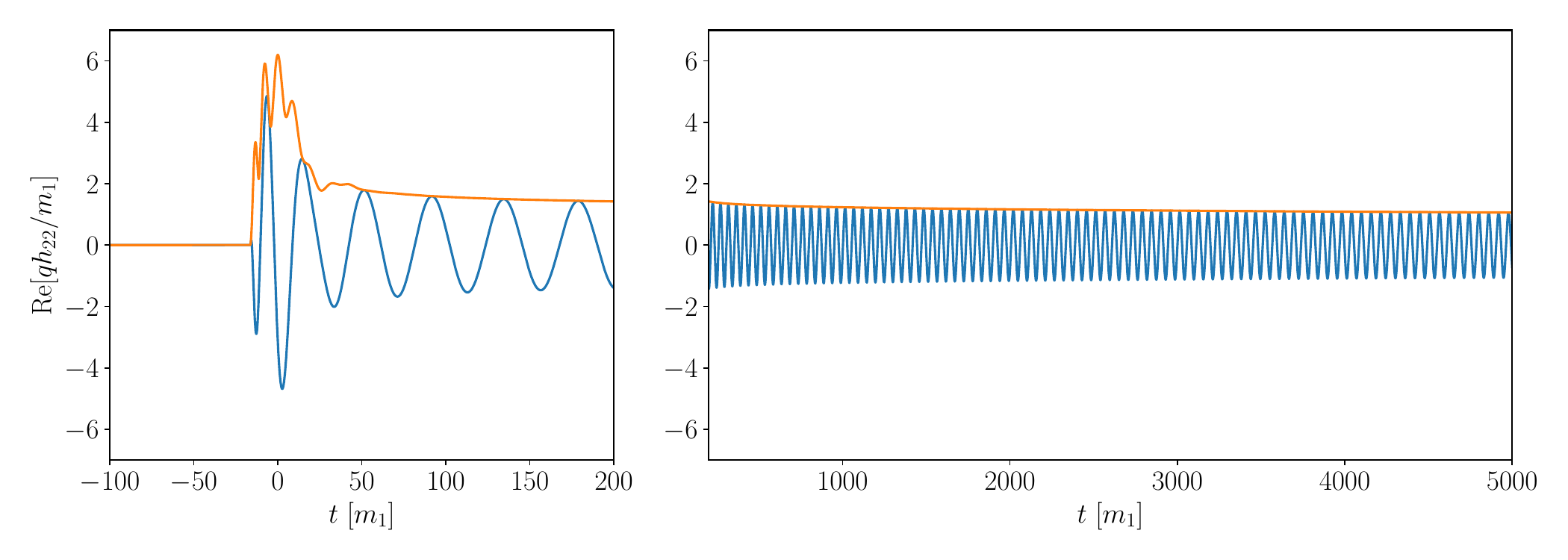}
\caption{\textit{Black-hole emission in radiation-reaction driven orbits:} The real part of the $(2,2)$ mode waveform - in a black hole emission scenario where $q=10^3$ - is presented by the blue solid line. Additionally, the amplitudes are shown in orange. More details can be found in Section \ref{sec:discussion}.}
\label{fig:radiation_reaction_waveforms}
\end{figure*}

\begin{figure*}[h]
\includegraphics[width=\textwidth]{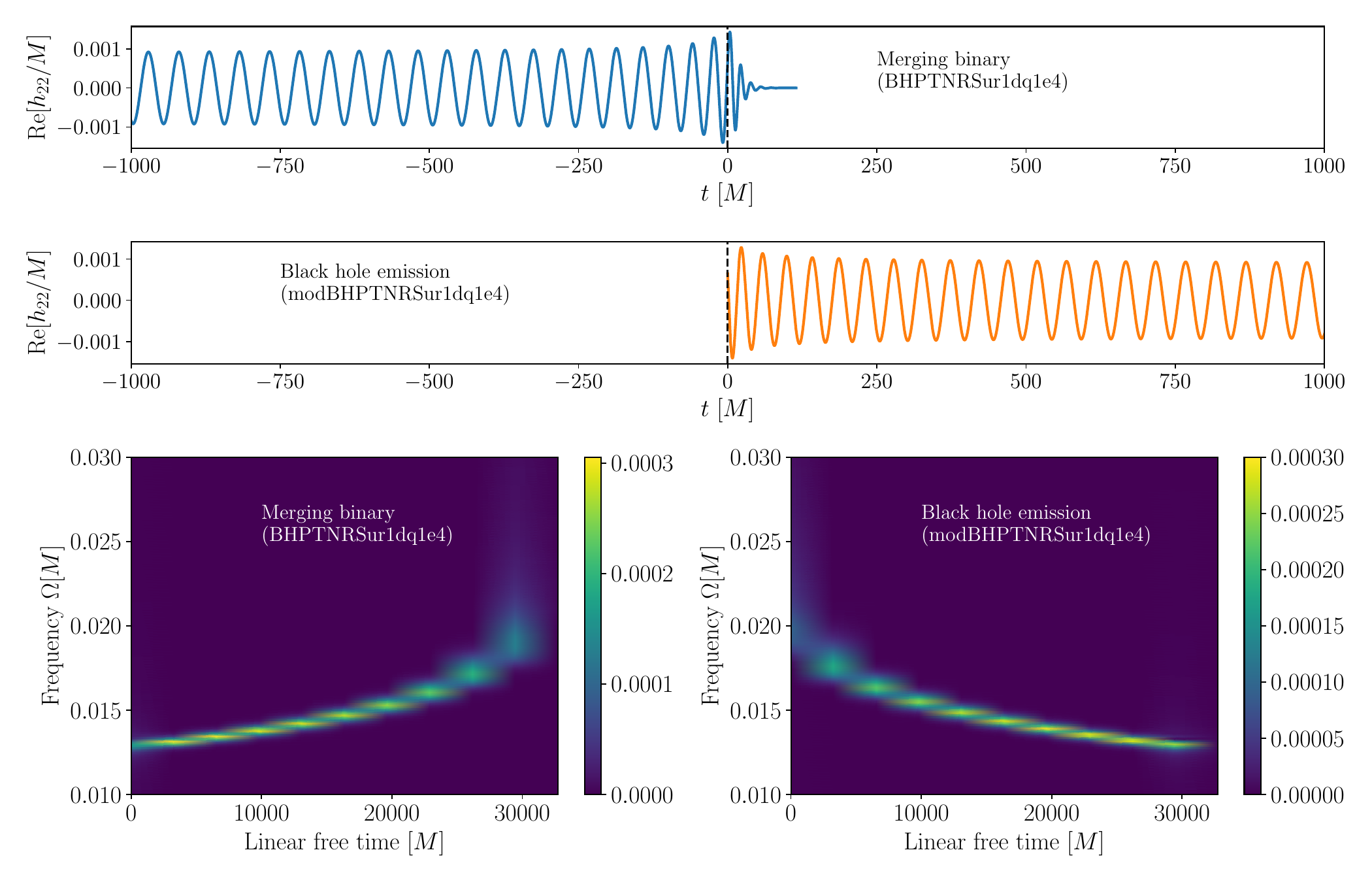}
\caption{\textit{Spectograms:} \textbf{Top:} The real part of the $(2,2)$ mode waveform (obtained using the \texttt{BHPTNRSur1dq1e4} model) for a merging binary with $q=1000$.
\textbf{Middle:} The real part of the $(2,2)$ mode waveform produced by a black hole emission event (obtained by modifying the \texttt{BHPTNRSur1dq1e4} waveform).
\textbf{Bottom:} Spectrogram of the gravitational waveform associated with the binary black hole merger (left) and black hole emission (right). More details can be found in Section \ref{sec:discussion}.}
\label{fig:spectrogram}
\end{figure*}

Once the conventional inspiral-plunge orbit is obtained, we reverse this complete trajectory to construct our hypothetical black hole emission orbit. Of course, it is worth noting that reversing an inspiral trajectory i.e. an ``outspiral'' doesn't have strong physical justification. We proceed with this approach nonetheless since we are not confining ourselves to standard classical general relativistic physics. This reversed trajectory is then input into the time-domain Teukolsky solver code mentioned in Section~\ref{sec:method}, and the resulting gravitational waveform is computed. In Figure~\ref{fig:radiation_reaction_waveforms}, we present the real part of the $(2,2)$ mode of the radiation waveform as a blue solid line. Additionally, for visual inspection, we overlay the amplitude on top of the real part of the $(2,2)$ mode. 

We find that even for a radiation-reaction driven orbit, the gravitational waveform obtained from black hole emission looks like a reverse chirp signal. For the waveform (and amplitude) at the beginning of the emission signal (around the maximum amplitude), oscillatory features are observed, possibly related to junk radiation, similar to the geodesic cases. Furthermore, like the geodesic cases, these oscillations disappear approximately $50m_1$ after the maximum amplitude. We further confirm that this qualitative behavior remains the same for the higher-order modes as well.

\subsection{Waveform model}
While our understanding of the black hole emission and absorption problem is incomplete and in its infancy, the results presented thus far provide a straightforward method to generate hypothetical black hole emission and absorption waveforms using existing waveform models. The process involves: (i) generating a standard merging gravitational waveform, (ii) discarding the signal after the merger, and (iii) flipping the waveform in time (and appending it to a standard merging waveform in case of an absorption event afterward). 

To illustrate this procedure, we modify \texttt{BHPTNRSur1dq1e4}~\cite{Islam:2022laz}, a state-of-the-art waveform model for merging binary black holes, using the \texttt{BHPTNRSurrogate} package available in the Black Hole Perturbation Toolkit \cite{BHPToolkit}. We develop a small module that takes the original \texttt{BHPTNRSur1dq1e4} waveform and adapts it to mimic waveforms emitted in a black hole emission and absorption scenario. This module will be available through the \texttt{gwModels} waveform package.

In Figure~\ref{fig:spectrogram} (upper panel), we show the $(2,2)$ mode waveforms obtained using the \texttt{BHPTNRSur1dq1e4} model for a merging binary with $q=1000$. We then modify the waveform to obtain a hypothetical black hole emission waveform (Figure~\ref{fig:spectrogram}; lower panel). 

\subsection{Detection}
Finally, we show the spectrogram of the gravitational waveforms associated with the binary black hole merger and black hole emission in the lower panel. To obtain the spectrogram, we compute the short time Fourier transform of the waveform. This is done using \texttt{scipy.signal.stft} module. We observe that the spectrogram clearly depicts a chirp signal (where the frequency increases with time) for the binary black hole merger, whereas a reverse chirp signal (where the frequency decreases monotonically with time) is evident for the black hole emission scenario. This striking difference could serve as an initial indicator of a black hole emission signal in gravitational wave detector data. 

\begin{figure}
\includegraphics[width=\columnwidth]{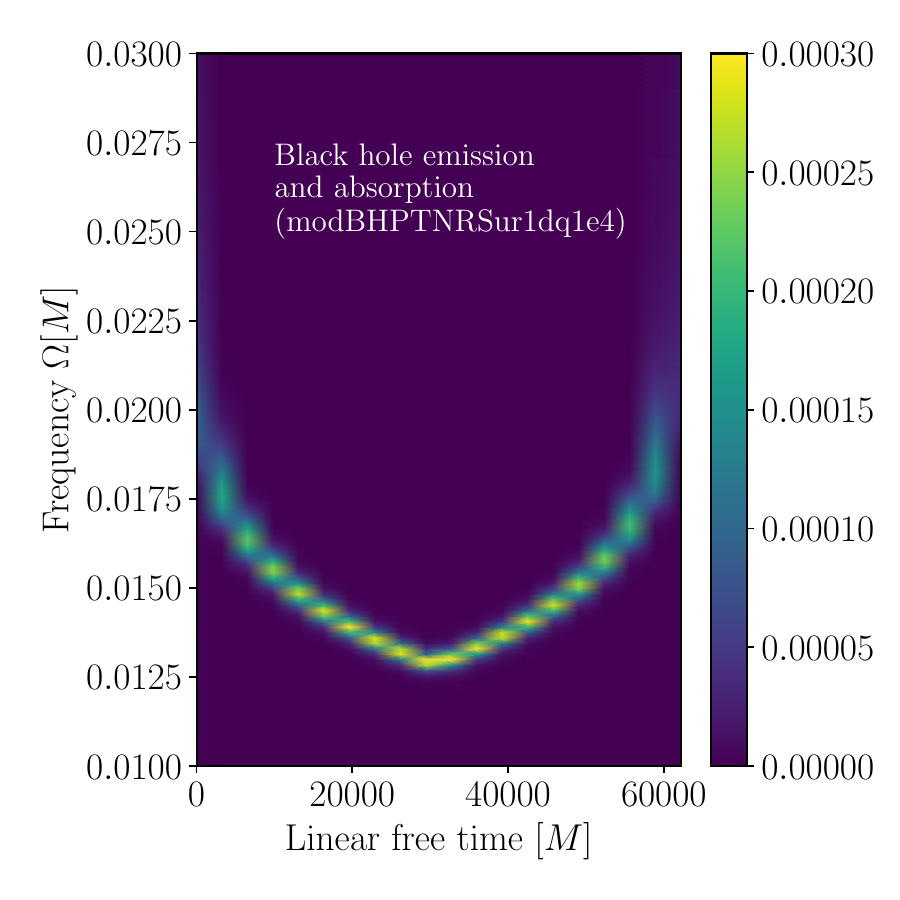}
\caption{\textit{Spectrogram:}  Spectrogram of the gravitational waveform associated with the binary black hole merger (left) and black hole emission (right). More details can be found in Section \ref{sec:discussion}.}
\label{fig:spectrogram_emission_absorption}
\end{figure}

The spectrogram for the black hole emission and subsequent absorption scenario (Figure~\ref{fig:spectrogram_emission_absorption}) exhibits an intriguing feature. It displays an initial reverse chirp (corresponding to the black hole emission) followed by a chirp (corresponding to the black hole absorption). The overall signal looks like a set of antlers in the spectrogram. 

While the presence of reverse chirp signals or antler-like features in the strain data recorded by the LIGO-Virgo-KAGRA collaboration~\cite{KAGRA:2020tym,VIRGO:2014yos,LIGOScientific:2014pky} would be an initial indication of possible black hole emission (and subsequent absorption), conducting a fully consistent detection analysis would first require constructing a template bank for the search. Such a bank could, in principle, be created using simple modifications to existing waveform models, like the one presented here. However, constructing a template bank for gravitational wave detection is complex, and its difficulty increases non-trivially with the dimensionality (i.e., the degrees of freedom in the binary parameter space). For scenarios involving only absorption (and no emission), the emitted waveform will be quite similar to conventional merging BBH waveforms. Therefore, existing search pipelines are expected to find such signals without requiring modifications. However, distinguishing between conventional BBH waveforms and absorption signals remains unclear, as they match quite well. We therefore leave these tasks for the future. 

\section{Concluding remarks}
\label{sec:conclusion}
Motivated by ideas from high-energy theory such as the weak gravity conjecture, white holes, wormholes, and Hawking radiation, we simulate the gravitational waveform resulting from the emission of a small black hole by a larger primary black hole using adiabatic point-particle black-hole perturbation theory. We first obtain the trajectory for the secondary by reversing the geodesic plunge. From this reversed trajectory, we solve the Teukolsky equation for the emission waveform. Similarly, by combining a reversed trajectory with a conventional geodesic plunge, we construct the trajectory for the black hole emission and subsequent absorption. We then use this trajectory to obtain the waveform.

To compare this waveform with the geodesic plunge waveform, we flip the emission waveform in time. We find evidence that at least a significant portion of the difference between these two waveforms (particularly around the merger) arises from resolution-dependent effects associated with junk radiation. An important conclusion from our work is that the emission waveform should be expected to resemble the time reversal of a plunge waveform. Exploiting the phenomenology of black hole emission and absorption signals, we then introduce straightforward modifications to existing gravitational waveform models to mimic gravitational radiation associated with these exotic events. We clearly demonstrate that while a typical binary signal exhibits a chirp behavior in the spectrogram, a black hole emission (and absorption) signal displays a distinctive reverse chirp behavior (antler-like feature) in the spectrogram. 
This reverse chirp characteristic should provide for easy detection.
We anticipate that these initial simulations, albeit incomplete, along with the adjusted waveform models, will contribute to the development of null tests for GR with GWs.

\section*{Acknowledgments} 
The authors acknowledge the support and hospitality of the Institute for Computational and Experimental Research in Mathematics in Providence, RI where this work began, and we 
thank Carl-Johan Haster for helpful discussions while there. Part of this work is additionally supported by the Heising-Simons Foundation, the Simons Foundation, and NSF Grants Nos. PHY-1748958. G.K. acknowledges support from NSF grants No. DMS-2309609 and PHY-2307236. 
S.L. acknowledges support from NSF grants PHY-2308861 and PHY-2409407 and computing resources from ACCESS and TACC's \texttt{frontera}.
Simulations were performed on CARNiE at the Center for Scientific Computing and Visualization Research (CSCVR) of UMassD, which is supported by the ONR/DURIP Grant No.\ N00014181255 and the UMass-URI UNITY HPC/AI supercomputer supported by the Massachusetts Green High Performance Computing Center (MGHPCC).

\bibliographystyle{utphys}
\bibliography{bib}
	
\end{document}